\documentclass[journal]{IEEEtran}

\IEEEoverridecommandlockouts

\usepackage{mathrsfs}
\usepackage{latexsym}
\usepackage{graphicx}
\usepackage{epsfig}
\usepackage{subfigure}
\usepackage{array}
\usepackage{amsmath}
\usepackage{amssymb}
\usepackage{color, soul}
\usepackage{amsthm}
\usepackage{enumerate}
\usepackage{algpseudocode}
\usepackage{algorithm}
\usepackage{bm}
\usepackage{amssymb}
\usepackage{balance}

\theoremstyle{plain} 

\theoremstyle{definition}

\def\bal#1\eal{\begin{align}#1\end{align}}

\newcommand{\bxi} {\boldsymbol{\xi}}

\newcommand{\bH}{{\bf H}}
\newcommand{\bJ}{{\bf J}}
\newcommand{\bR}{{\bf R}}

\newcommand{\bI}{{\bf I}}

\newcommand{\bV}{{\bf V}}
\newcommand{\bv}{{\bf v}}

\newcommand{\bA}{{\bf A}}

\newcommand{\bB}{{\bf B}}

\newcommand{\bQ}{{\bf Q}}
\newcommand{\bL}{{\bf L}}
\newcommand{\bZ}{{\bf Z}}

\newcommand{\bq}{{\bf q}}
\newcommand{\bT}{{\bf T}}

\newcommand{\bX}{{\bf X}}

\newcommand{\gl}{\lambda}

\newcommand{\by}{{\bf y}}
\newcommand{\bx}{{\bf x}}

\newcommand{\ba}{{\bf a}}

\newcommand{\bo}{{\bf 0}}

\newcommand{\bp} {\begin{proof}}
\newcommand{\ep} {\end{proof}}

\newcommand{{\Rb}} {\right)}

\newcommand{{\Rf}} {\right\}}
\newcommand{{\diag}} {\mathrm{diag}}

\begin{document}

\title{Secure MIMO Transmission via Intelligent Reflecting Surface}

\author{Limeng Dong, Hui-Ming Wang \emph{Senior Member, IEEE}

\thanks{The work of H.-M. Wang was supported in part by the National Natural Science Foundation of China under Grant 61671364 and Grant 61941118, in part  by the Outstanding Young Research Fund of Shaanxi Province under Grant 2018JC 003, and in part by the Innovation Team Research Fund of Shaanxi Province under Grant 2019TD 013. (\emph{Corresponding author: Hui-Ming Wang.})}

\thanks{L. Dong and H.-M. Wang are with the  School of Information and Communications Engineering, Xi'an Jiaotong University, Xi'an, 710049, China, and also with the Ministry of Education Key Laboratory for Intelligent Networks and Network Security, Xi'an Jiaotong University, Xi'an 710049, China (e-mail: dlm$\_$nwpu@hotmail.com; xjbswhm@gmail.com)}

}

\maketitle

\begin{abstract}
 In this letter, we consider an intelligent reflecting surface (IRS) assisted Guassian multiple-input multiple-output (MIMO) wiretap channel in which a multi-antenna transmitter  communicates with a multi-antenna receiver in the presence of a multi-antenna eavesdropper. To maximize the secrecy rate of this channel,  an alternating optimization  (AO) algorithm  is proposed to jointly optimize the transmit covariance $\bR$ at transmitter and phase shift coefficient $\bQ$ at IRS by fixing the other as a constant. When $\bQ$ is fixed, existing numerical algorithm  is used to search for global optimal $\bR$. When $\bR$ is fixed, three sucessive approximation to the objective function  to surrogate lower bound is applied and minorization-maximization (MM) algorithm is proposed to optimize the local  optimal $\bQ$. Simulation results  have be provided to validate the convergence and performance of the proposed AO algorithm.
\end{abstract}	

\begin{IEEEkeywords}
Intelligent reflecting surface,  MIMO, phase shift, alternating optimization, MM.
\end{IEEEkeywords}

\section{Introduction}

 Intelligent reflecting surface (IRS) has
drawn wide attention for its applications in wireless communications. IRS is a  low complexity software-controlled passive metasurface which could significantly help enhancing user's transmission rate with very low power consumption \cite{Hu-18}.    Motivated by these advantages, IRS was recently applied to the study in physical layer security, and several research results about secrecy rate maximization of IRS-assisted  multiple-input multiple-output (MISO) wiretap channels wiretap channels were established, including single user case \cite{Shen-19}-\cite{Yu-19} and downlink multi-user case \cite{Chen-19}\cite{Xu-19}.   All these studies indicate that IRS significantly enhance user's secrecy rate.

However, all the aformentioned contributions in the current literatures \cite{Shen-19}-\cite{Xu-19} are only restricted to MISO  settings, i.e., only single antenna at the receiver as well as eavesdropper are considered. When  multi-input multi-output (MIMO) is considered, there are two significant differences about the optimization problems compared with that in conventional MISO case. Firstly, in MIMO systems, beamforming is not always optimal solution. Therefore, we need to optimize an transmit covariance instead of beamformer vector in the secrecy rate maximization problem for this case. Secondly, for secrecy rate optimization problems in the MIMO case, the objective function is  a complicated log of determinant formular compared with simple log of scalar formular for  the MISO case. Therefore, all these existing solutions for the MISO case fail to the MIMO case. To the best of our knowledge, the study of IRS-assisted MIMO wiretap channel is still an open problem and there is no existing numerical or analytical solutions to maximize its secrecy rate. 

Motivated by the aformetioned aspects, the main contribution in this letter is that we consider an IRS-assisted Gaussian MIMO wiretap channel, and aim at maximizing the user's secrecy rate numerically.  To solve this non-convex problem, an   alternating optimization  (AO) algorithm is proposed to jointly optimize the transmit covariance $\bR$ at transmitter as well as phase shift coefficient $\bQ$ at IRS. When $\bQ$ is fixed,  the existing algorithm is applied to optimize $\bR$ globally. When $\bR$ is fixed, we  approximate the  objective function to surrogate lower bound, and  the local optimal $\bQ$ is optimized via  minorization-maximization (MM) algorithm.  In particular, the key difficulty is how to find a proper surrogate function for the complicated objective function. Hence,  we apply three successive approximations for the objective function to obtain the proper lower bound so that MM alogrithm can be applied, which is significantly different from the existing MM used in the simple MISO case \cite{Shen-19}\cite{Yu-19} in which only one time approximation for the objective is needed due to the simple structure of the objective function.
As the convergence is reached, the  results returned by the AO algorithm is guaranteed to be a Karush-Khun-Tucker (KKT) solution of the original problem.

\emph{Notations}: $\bA^{T}$ and $\bA^{H}$ denote transpose  and  Hermitian conjugate of $\bA$, respectively; $\gl_{max}(\bA)$ denotes the maximum eigenvalues of $\bA$;  $|\bA|$ and $tr(\bA)$ are determinant and trace of $\bA$;  $\odot$ denotes Hadamard product; $arg(a)$ denotes the phase of the complex value $a$; $\bA_{ij}$ denotes the element in $i$-th row $j$-th column of $\bA$; $\ba_i$ is the $i$-th element of $\ba$.

\section{Channel Model And Problem Formulation}

Let us consider an IRS-assisted MIMO wiretap channel model shown as Fig. 1, in which a  transmitter Alice, receiver Bob, eavesdropper Eve and an IRS are included. The  number of antennas deployed at Alice, Bob and Eve are $m$, $d$, $e$ respectively, and the number of reflecting elements on the IRS is $n$. We assume that Alice, Bob and Eve are located in city's hot spot area, and the direct link between Alice and Bob/Eve is blocked by a building. Then, the IRS is located in a higher position to help  Alice's  transmission by  passively reflecting the signals to Bob. Due to broadcast nautre of wireless channels, the reflected signal could also be sent to Eve. Therefore, the main task for IRS is to adjust the phase shift for signals by the reflecting elements  so as to increase the information rate at Bob but decrease the information leakage to Eve. Based on these settings, the received signals at Bob and Eve are expressed as
\bal
\label{channel}
\by_B = \bH_{IB}\bQ\bH_{AI}\bx+\bxi_{B},   \       \by_E = \bH_{IE}\bQ\bH_{AI}\bx+\bxi_{E},
\eal
respectively where $\bx$ is the transmitted signal, $\bQ$ is the diagonal phase shift matrix for IRS, in which the diagonal element is $e^{j\theta_i}$ ($i=1,2,...,n$), $\theta_i$ is the phase shift coefficient at reflecting element $i$, $\bH_{AI}$, $\bH_{IB}$ and $\bH_{IE}$  are the channel matrices representing the direct link of Alice-IRS, IRS-Bob and IRS-Eve respectively, $\bxi_{B}$ and $\bxi_{E}$ represent noise at  Bob and Eve respectively with i.i.d. entries distributed as $\mathcal{CN}(0,1)$.  And we consider that  full channel state information (CSI) is available at Alice, which can be achieved by modern adaptive system design, where channels are estimated at Bob and Eve, and send back to Alice. We note that Eve is just other
user in the system and it also share its CSI with Alice but is untrusted by Bob. The controller is used to coordinates Alice and IRS for channel acquisition and data transmission tasks \cite{Cui-19}.

\begin{figure}[t]
	\centerline{\includegraphics[width=3.0in]{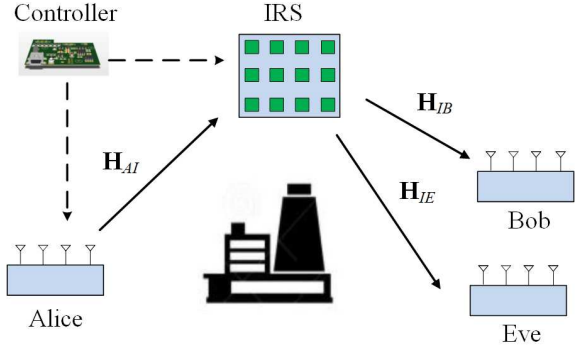}}
	\caption{A block diagram of IRS-assisted Gaussian MIMO wiretap channel}
\end{figure}

Given \eqref{channel}, the secrecy rate maximization of this model can be expressed as the following problem P1.
\bal
\notag
P1: \underset{\bR, \bQ}{\max}\ C_s(\bR,\bQ)=log_2\frac{|\bI+\bH_B\bR\bH_B^H|}{|\bI+\bH_E\bR\bH_E^H|}\\
s.t.\ \ \bR\geq \bo, tr(\bR)\leq P, |\bQ_{i,i}|=1, (i=1,2,...,n)
\eal
where $\bH_i=\bH_{Ii}\bQ\bH_{AI}, i\in\{B, E\}$, $P$ denotes  total transmit power budget for Alice, $\bR=E\{\bx\bx^H\}$ stands for the transmit covariance for Alice and where the unit modulus constraint $|\bQ_{i,i}|=1$ ensures that each reflecting element in IRS does not change the amplitude of the signals.  Obviously, the determinant part in the objective function of P1 cannot be simplified to scalar formular  as for MISO case, thereby significantly increasing the difficulty to solve this problem.

\section{Alternating Optimization Algorithm}
To solve this new non-convex problem,  we propose an iterative AO algorithm  to   optimize $\bR$ and $\bQ$   alternatively by fixing the other as constant. 

We firstly fix  $\bQ$ as a constant and maximize $\bR$. Note that when  $\bQ$ is fixed, $\bH_B$ and $\bH_E$ are also fixed so that P1 is reduced to a secrecy capacity optimization problem of general Gaussian MIMO wiretap channel. To solve this problem, we apply the key Theorem 1 in \cite{Loyka-15} so that the original problem is equivalently transformed to a convex-concave max-min optimization problem. Then, we apply the existing algorithm  in \cite{Loyka-15} which is based on barrier method in combination with Newton method and backtracking line search method to globally optimized $\bR$. Note that in \cite{Loyka-15}, the algorithm was developed only based on  real-valued channel matrix case. Therefore, we extend this algorithm to   complex-valued channel  cases by re-deriving the gradient and Hessians of the barrier objective function. Using the same steps of proof illustrated in \cite{Loyka-15}, the non-singularity of Hessian matrix as well as global convergence of the algorithm still can be proved for the complex-valued channel case. Here we omit the  detailed steps of this algorithm due to page limit.  

The next step is to optimize $\bQ$ for fixed $\bR$ in the subproblem, which can be express   as P2.
\bal
\notag
P2: &\underset{\bQ}{\max}\ f(\bQ)=f_B(\bQ)+f_E(\bQ)\\
\label{unit}
 &s.t.\ \  |\bQ_{i,i}|=1,  i=1,2,...,n
\eal
where $f_B(\bQ)=log_2|\bI+\bH_{IB}\bQ\bL\bQ^H\bH_{IB}^H|$, $f_E(\bQ)=-log_2|\bI+\bH_{IE}\bQ\bL\bQ^H\bH_{IE}^H|$ 
and where  $\bL=\bH_{AI}\bR\bH_{AI}^H$. P2 is a complicated non-convex problem with both non-convex objective function and constraints, and the existing solutions (e.g., semi-definite relaxation and fractional programming) for IRS-assisted MISO case \cite{Shen-19}\cite{Cui-19} cannot be directly applied to our problem. To optimize $\bQ$, we apply MM algorithm to solve P2, in which the key idea  is to firstly obtain an approximately lower  bound (i.e., surrogate function) of the objective, and then iteratively compute the optimal value of this bound subject to the constraints. If the bound is constructed properly, any converged point  genrated by MM is a KKT point (i.e., local optimal point) for the original problem. For detailed explanation of MM, please refer to \cite{Sun-17}. 

 Since the objective function $f(\bQ)$ in P2  consists of two complicated log determinant functions,  the difficulty to directly find its  proper lower bound has significantly increased. Therefore, the solution we apply is firstly find the lower bound for $f_B(\bQ)$ and $f_E(\bQ)$ respectively, and then formulate the new approximated problem by  adding this two bounds together. After that, we make further two successive approximations to this bound to formulate the final surrogate function of $f(\bQ)$, and apply MM algorithm to optimize a local optimal solution of $\bQ$.

Firstly,  consider $\tilde{\bQ}$ is a feasible point satisfying the unit mudulous constraint, a quadratic lower bound of the function $f_E(\bQ)$ can be expressed as
\bal
\label{upper_f_E}
f_E(\bQ)\geq\bar{f}_E(\bQ,\tilde{\bQ})+C_1(\tilde{\bQ})
\eal
where 
\bal
\notag
&\bar{f}_E(\bQ,\tilde{\bQ})=-tr(\tilde{\bQ}_E^{-1}\bH_{IE}\bQ\bL\bQ^H\bH_{IE}^H),\\ 
\notag
&C_1(\tilde{\bQ})=-log_2|\tilde{\bQ}_E|+tr(\tilde{\bQ}_E^{-1}\bH_{IE}\tilde{\bQ}\bL\tilde{\bQ}^H\bH_{IE}^H),\\ 
\notag
&\tilde{\bQ}_E=\bI+\bH_{IE}\tilde{\bQ}\bL\tilde{\bQ}^H\bH_{IE}^H. 
\eal
The inequality in \eqref{upper_f_E} is obtained via the lemma: for any matrix $\bA\in\mathbb{C}^{n\times n}$ and $\tilde{\bA}\in\mathbb{C}^{n\times n}$, 
\bal
\label{lemma_logA}
log_2|\bA| \leq log_2|\tilde{\bA}|+tr(\tilde{\bA}^{-1}(\bA-\tilde{\bA})).
\eal
Then, to obtain the lower bound of $f_B(\bQ)$, let $\bT=\bH_{IB}\bQ\bL^{\frac{1}{2}}$, according to matrix inversion lemma, $f_B(\bQ)$ can be further expressed as
\bal
f_B(\bQ)=-log_2|\bI-\bT(\bI+\bT^H\bT)^{-1}\bT^H|.
\eal
Let $\tilde{\bT}=\bH_{IB}\tilde{\bQ}\bL^{\frac{1}{2}}$, by applying \eqref{lemma_logA}, $f_B(\bQ)$ is also lower bounded by
\bal
\notag
f_B(\bQ)&\geq -log_2|\tilde{\bQ}_B|-tr(\tilde{\bQ}_B^{-1}(\bQ_B-\tilde{\bQ}_B))\\
&=C_2(\tilde{\bQ})+h_B(\bQ)
\eal
where  $C_2(\tilde{\bQ})=-log_2|\tilde{\bQ}_B|+tr(\bI)-tr(\tilde{\bQ}_B^{-1})$, $h_B(\bQ)=tr(\tilde{\bQ}_B^{-1}\bT(\bI+\bT^H\bT)^{-1}\bT^H)$, $\tilde{\bQ}_B=\bI+\bH_{IB}\tilde{\bQ}\bL\tilde{\bQ}^H\bH_{IB}^H$.

Hence, combining (4) and (7), the approximated problem of P2 can be expressed as P3
\bal
\notag
P3: \max_{\bQ} \ \bar{f}_E(\bQ,\tilde{\bQ})+C_1(\tilde{\bQ})+C_2(\tilde{\bQ})+h_B(\bQ),  s.t. \ \eqref{unit}.
\eal
However,  we find that it is still difficult to optimize $\bQ$ given $\tilde{\bQ}$ due to the complicated  term $h_B(\bQ)$ as well as the constraint \eqref{unit}. Therefore, a second approximation of the objective function in P3 is needed. In the following, we apply the key lemma of matrix fractional functions \cite{Boyd-04}: for any positive semi-definite matrix $\bA\in\mathbb{C}^{m \times m}$ and positive definite matrix $\bB, \tilde{\bB}\in\mathbb{C}^{n \times n}$, and $\bX, \tilde{\bX}\in\mathbb{C}^{m \times n}$, 
\bal
\notag
&tr(\bA\bX\bB^{-1}\bX^H)\\
\notag
&\geq tr(\bA\tilde{\bX}\tilde{\bB}^{-1}\tilde{\bX}^H)-tr(\bA\tilde{\bX}\tilde{\bB}^{-1}(\bB-\tilde{\bB})\tilde{\bB}^{-1}\tilde{\bX}^H)\\
\notag
&+tr(\bA(\bX-\tilde{\bX})\tilde{\bB}^{-1}\tilde{\bX}^H)+tr(\bA\tilde{\bX}\tilde{\bB}^{-1}(\bX-\tilde{\bX})^H).
\eal
Therefore, by applying this lemma to the term $h_B(\bQ)$  via setting $\bA=\tilde{\bQ}_B^{-1}$, $\bX=\bT$, $\tilde{\bX}=\tilde{\bT}$, $\bB=\bI+\bT^H\bT$ and $\tilde{\bB}=\bI+\tilde{\bT}^H\tilde{\bT}$ and after some manipulations, the lower bound of $f(\bQ)$ can be further expressed as
\bal
\notag
&f(\bQ)\geq\bar{f}_E(\bQ,\tilde{\bQ})+C_1(\tilde{\bQ})+C_2(\tilde{\bQ})+h_B(\bQ)\\
&\geq\bar{f}_E(\bQ,\tilde{\bQ})+C_1(\tilde{\bQ})+C_2(\tilde{\bQ})+g_B(\bQ)+C_3(\tilde{\bQ})
\eal
where
\bal
\notag
&C_3(\tilde{\bQ})=-tr(\tilde{\bQ}_B^{-1})+tr(\tilde{\bQ}_B^{-1}\bJ_B\tilde{\bT}^H\tilde{\bT}\bJ_B^H),\\ 
\notag
&g_B(\bQ)=-tr(\tilde{\bQ}_B^{-1}\bJ_B\bT^H\bT\bJ_B^H)+tr(\tilde{\bQ}_B^{-1}\bJ_B\bT^H)\\&\ \ \ \ \ \ \ \ \ \ \ 
\notag
+tr(\bT\bJ_B^H\tilde{\bQ}_B^{-1})
\eal
and where $\bJ_B=\tilde{\bT}(\bI+\tilde{\bT}^H\tilde{\bT})^{-1}$. 

In the following, we express (8) to a more tractable form.  Let $\bq=[e^{j\theta_1},e^{j\theta_2},...,e^{j\theta_n}]^T$, and let
\bal
\notag
\bA_1=\tilde{\bQ}_B^{-1}\bJ_B\bL^{\frac{1}{2}},\ \bA_2=\bH_{IB}^H\bH_{IB},\ \bA_3=\bL^{\frac{1}{2}}\bJ_B^H,\\
\notag
     \bA_4=\bH_{IB}^H\tilde{\bQ}_B^{-1}\bJ_B\bL^{\frac{1}{2}},\ \bA_5=\tilde{\bQ}_E^{-1}\bH_{IE}.
\eal
We firstly apply the lemma of matrix identity in \cite{Zhang-17}: for any matrix $\bA$, $\bB$ and diagonal matrix $\bV$ with proper sizes, $tr(\bV^H\bA\bV\bB)=\bv^H(\bA\odot\bB^T)\bv$ holds where the entries in $\bv$ are all diagonal elements in $\bV$. Using this lemma, $f(\bQ)$ is lower bounded as
\bal
\notag
f(\bQ)\geq&-tr(\bQ^H\bH_{IE}^H\bA_5\bQ\bL)-tr(\bQ^H\bA_2\bQ\bA_3\bA_1)\\
\notag
&+tr(\bA_4\bQ^H)+tr(\bQ\bA_4^H)+\sum_{i=1}^{3}C_i(\tilde{\bQ})\\
=&-g(\bq)+2Re\{\bq^H\ba\}+\sum_{i=1}^{3}C_i(\tilde{\bQ})
\eal
where 
$g(\bq)=\bq^H\bZ\bq$, $\bZ=\bA_2\odot(\bA_3\bA_1)^T+(\bH_{IE}^H\bA_5)\odot\bL^T$ 
and where the entries in $\ba$ are all diagonal entires in $\bA_4$. It can be known that given fixed feasible $\tilde{\bQ}$, the bound (9) is a quadratic convex function respect to $\bq$. However, since $\bq$ needs to  satisfy  the non-convex unit modulus constraint $|\bq_i|=1$, it is still difficult to use MM algorithm to optimize (9). Hence, a third approximation of $f(\bQ)$ is needed by finding a surrogate function of $g(\bq)$, which is expressed as follows \cite{Sun-17}.
\bal
\notag
g(\bq) \leq &\bq^H\gl_{max}(\bZ)\bI\bq-2Re\{\bq^H(\gl_{max}(\bZ)\bI-\bZ)\tilde{\bq}\}\\
\notag
&+\tilde{\bq}^H(\gl_{max}(\bZ)\bI-\bZ)\tilde{\bq}\\
\notag
=&2n\gl_{max}(\bZ)-2Re\{\bq^H(\gl_{max}(\bZ)\bI-\bZ)\tilde{\bq}\}-\tilde{\bq}^H\bZ\tilde{\bq}\\
=&\tilde{g}(\bq,\tilde{\bq})
\eal
where $\tilde{\bq}$ is the feasible point, the entries of which are the diagonal entries of $\tilde{\bQ}$. Hence, $f(\bQ)$ can be further approximated by
\bal
f(\bQ)\geq -\tilde{g}(\bq,\tilde{\bq})+2Re\{\bq^H\ba\}+\sum_{i=1}^{3}C_i(\tilde{\bQ})
\eal
By dropping the constant terms of this bound, P3 can be further approximated to P4.
\bal
\notag
P4:\ \max_{\bq}\ \ Re\{\bq^H\bv\}\ \ \ s.t.\ |\bq_i|=1, i=1,2,...,n
\eal
where $\bv=(\gl_{max}(\bZ)\bI-\bZ)\tilde{\bq}+\ba$. Obviously, the objective is maximized only when the phase of $\bq_i$ and $\bv_i$ are equal. Thus, the closed-form global optimal solution for P4 is expressed as
\bal
\label{optimal_q}
\bq_{opt}=[e^{jarg(\bv_1)}, e^{jarg(\bv_2)},..., e^{jarg(\bv_n)}]^T.
\eal
Therefore, by initializing a feasible point $\tilde{\bq}$ and use MM to optimize P4,  a KKT  solution of P2 given fixed $\bR$ can be obtained.

Based on the above analysis,  the overall AO algorithm for maximizing  the secrecy rate of IRS-assisted MIMO wiretap channel is summarized as Algorithm 1. Since $\bR$  and $\bQ$ are optimized alternatively, the objective function $C_s(\bR_k,\bQ_k)$ is monotonically increasing with number of iterations $k$. Moreover, since $\bR$  and $\bQ$ are both bounded by the independent constraints in P1, according to Cauchy's theorem \cite{Chen-19}, the algorithm is guaranteed to converge.

\begin{algorithm}[h]
	\caption{(\it AO algorithm of solving P1)}
	\begin{algorithmic}
    \Require Starting point  $\bR_0$ and $\bQ_0$.
        \State 1. Set  $k=0$, compute $C_s(\bR_0,\bQ_0)$.
		\Repeat\ \  (AO algorithm) 
         \State 2. Set $k=k+1$, optimize $\bR_k$ given fixed $\bQ_{k-1}$ via barrier method in \cite{Loyka-15}. 
          \State 3. Optimize $\bQ_k$ given fixed $\bR_k$ via MM algorithm.
           \State 4.  Compute $C_s(\bR_k,\bQ_k)$.
    \Until $|C_s(\bR_k,\bQ_k)-C_s(\bR_{k-1},\bQ_{k-1})|/|C_s(\bR_{k-1},\bQ_{k-1})|$ converges to certain accuracy
       \State 5. Output $\bR_k$, $\bQ_k$ as  KKT point of P1.
	\end{algorithmic}
\end{algorithm}

\section{Simulation Results}
To validate the convergence and peformance of our proposed AO algorithm,  simulation results have been carried out in this section. We consider a fading environment, and all the channels $\bH_{AI}$, $\bH_{IB}$ and $\bH_{IE}$ are  formulated as the product of large scale fading and small scale fading. The entries in the small scale fading matrix are   randomly generated  with complex zero-mean
Gaussian random variables with unit covariance. For the large scale fading in all links of Alice-IRS, IRS-Bob and IRS-Eve, we refer to \cite{Cui-19} by setting the path loss as -30dB at reference distance 1m, and path loss exponents as 3. In AO algorithm, we set the target accuracy for barrier method as $10^{-8}$ and for MM algorithm as $10^{-4}$.

 Fig. 2 illustrates the convergence of the objective $C_s(\bR_k,\bQ_k)$ as function of number of iterations $k$ in the proposed AO algorithm under randomly generated channels with different settings of $m,d,e$ and $n$. Based on the results, it requires 42, 107 and 166 steps for $C_s(\bR_k,\bQ_k)$ to converge to $10^{-4}$ for  each considered setting, also note that the convergence  is monotonically increasing. In fact, given fixed target accuracy, larger settings of $m,n$ leads to larger dimensions of variable $\bR$ and $\bQ$ so that the AO algorithm requires more iterations to optimize each element of these variables. In addition to these results, our extensive simulations show a monotonic convergence of AO algorithm.

\begin{figure}[t]
	\centerline{\includegraphics[width=3.0in]{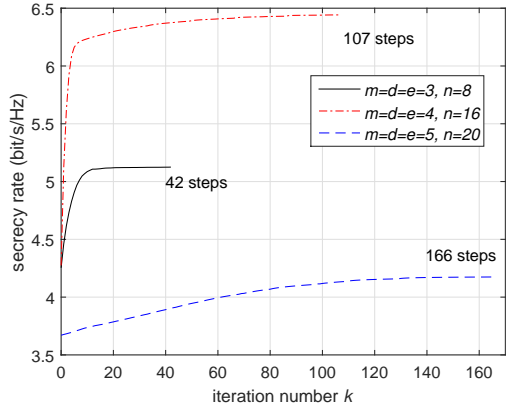}}
	\caption{Convergence of $C_s(\bR_k,\bQ_k)$ under different settings of $m,n,d,e$, $P$ is fixed at 35dBm. The convergence is monotonic for all considered cases.}
\end{figure}

In Fig. 3, we compare the performance of our AO algorithm with two benchmark schemes: 1)optimize $\bR$ given zero phase shift (i.e., $\bQ=\bI$) at IRS; 2)optimize $\bR$ given random phase shift at IRS. The results are averaged over 100 randomly generated channels. According to the figure, we note that our proposed AO algorithm has significantly better performance than the other two benchmark schemes. For both zero phase shift and random phase shift methods, it can be seen that only optimizing $\bR$ at transmitter has very limited performance on enhancing secrecy rate. For random phase shift method, the randomly generated $\bQ$ can deteriorate quality of effective channel $\bH_B$ but improving  $\bH_E$ in some channel realization cases so that it has least performance.

\begin{figure}[t]
	\centerline{\includegraphics[width=3.0in]{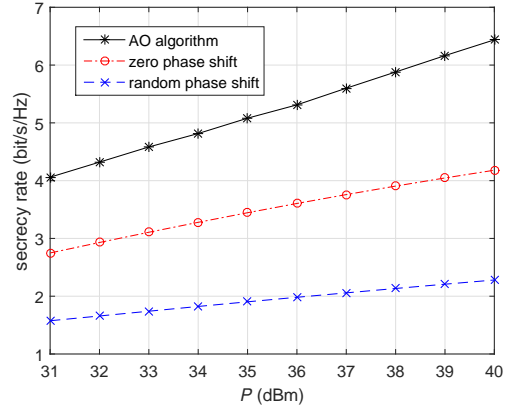}}
	\caption{ Performance comparison of achieved secrecy rate under  $m=d=e=5$ and $n=15$. The results are averaged over 100 randomly generated channels.}
\end{figure}

\section{Conclusion}
In this letter, the secrecy rate maximization problem of an IRS-assisted Gaussian MIMO wiretap channel is studied. To solve this difficult non-convex problem, an AO algorithm  is proposed to jointly optimize the transmit covariance  at Alice and phase shift coefficient at IRS.  Simulation results have validated the monotonic convergence of the proposed AO algorithm, and it is shown that the performance of AO is significantly better  than the other benchmark schemes.


\end{document}